\documentclass[acmsmall,screen]{acmart}

\AtBeginDocument{%
  }

\setcopyright{acmlicensed}
\copyrightyear{2018}
\acmYear{2018}
\acmDOI{XXXXXXX.XXXXXXX}

\acmConference[Conference acronym 'XX]{Make sure to enter the correct
  conference title from your rights confirmation emai}{June 03--05,
  2018}{Woodstock, NY}
\acmISBN{978-1-4503-XXXX-X/18/06}

\usepackage{algorithmic}
\usepackage{graphicx}
\usepackage{textcomp}
\usepackage{xcolor}

\usepackage{pifont}
\usepackage{xspace}
\usepackage{epigraph}
\usepackage{booktabs}
\usepackage{multirow}
\usepackage[most]{tcolorbox}
\usepackage{colortbl}
\usepackage{algorithm}
\usepackage{enumitem}
\usepackage{xspace}
\usepackage{wasysym}

\def\eg{\textit{e.g.,} }
\def\ie{\textit{i.e.,} }
\def\method{{\sc HonestCoder}\xspace}
\def\bench{{\sc TruthCodeBench}\xspace}

\newcommand{\greyboxb}[2]{
\vspace{0.05cm}
    \begin{tcolorbox}[
        left=2pt, right=2pt, top=2pt, bottom=2pt,
        boxrule=0.2mm,
        leftrule=2mm,
        arc=0mm,
        colframe=black!40!white,
        colback=black!5!white,
        colbacktitle=black!50!white
    ]
    \textbf{#1}{#2}
    \end{tcolorbox}
\vspace{0.05cm}
}

\begin{document}

\title{Showing LLM-Generated Code Selectively Based on Confidence of LLMs}

\author{Jia Li $\male$}
\email{lijia@stu.pku.edu.cn}
\orcid{0002-5579-8852}
\affiliation{%
  \institution{School of Computer Science, Peking University}
  \city{Beijing}
  \country{China}
}
\author{Yuqi Zhu}
\email{lijia@stu.pku.edu.cn}
\orcid{0009-0004-2903-7138}
\affiliation{%
  \institution{Academy of Military Sciences}
  \city{Beijing}
  \country{China}
}
\author{Yongmin Li}
\email{liyongmin@pku.edu.cn}
\orcid{0009-0001-3702-0043}
\author{Ge Li}
\email{lige@pku.edu.cn}
\orcid{0000-0002-5828-0186}
\author{Zhi Jin}
\email{zhijin@pku.edu.cn}
\orcid{0000-0003-1087-226X}
\affiliation{%
  \institution{School of Computer Science, Peking University}
  \city{Beijing}
  \country{China}
}

\begin{abstract}
Large Language Models (LLMs) have shown impressive abilities in code generation, but they may generate erroneous programs. Reading a program takes ten times longer than writing it \cite{Clean_Code}. Showing these erroneous programs to developers will waste developers' energies and introduce security risks to software.

To address the above limitations, we propose \method, a novel LLM-based code generation approach. \method selectively shows the generated programs to developers based on LLMs' confidence. The confidence provides valuable insights into the correctness of generated programs. To achieve this goal, we propose a novel approach to estimate LLMs' confidence in code generation. It estimates confidence by measuring the multi-modal similarity between LLMs-generated programs. 

We collect and release a multilingual benchmark named \bench, which consists of 2,265 samples and covers two popular programming languages (\ie Python and Java). We apply \method to four popular LLMs (\eg DeepSeek-Coder and Code Llama) and evaluate it on \bench. Based on the experiments, we obtain the following insights. 
\ding{182} \method can effectively estimate LLMs' confidence and accurately determine the correctness of generated programs. For example, \method outperforms the state-of-the-art baseline by 27.79\% in AUROC and 63.74\% in AUCPR. 
\ding{183} \method can decrease the number of erroneous programs shown to developers. Compared to eight baselines, it can show more correct programs and fewer erroneous programs to developers.
\ding{184} Compared to showing code indiscriminately, \method only adds slight time overhead (approximately 0.4 seconds per requirement).
\ding{185} We discuss future directions to facilitate the application of LLMs in software development. We hope this work can motivate broad discussions about measuring the reliability of LLMs' outputs in performing code-related tasks. 
\end{abstract}

\ccsdesc[300]{Computing methodologies~Neural networks}
\ccsdesc[500]{Software and its engineering~Automatic programming}

\keywords{Code Generation, Large Language Models}

\received{20 February 2007}
\received[revised]{12 March 2009}
\received[accepted]{5 June 2009}

\maketitle

\section{Introduction}
\label{sec:intro}

Large Language Models (LLMs) \cite{GPT-4o,DeepSeek-Coder,CodeLlama} have been widely used in code generation. For example, GitHub Copilot \cite{Copilot}, an LLM-based code generation tool, is regularly utilized by developers from over 10,000 organizations \cite{Copilot_Talk}. However, LLMs may produce erroneous programs \cite{Code_Error_Study1,Code_Error_Study2}. Showing these erroneous programs to developers leads to two negative consequences. \ding{182} \textbf{Wasting developers' energies.} Reading a program takes ten times longer than writing it \cite{Clean_Code}. Developers waste more time reading and testing erroneous programs than writing them. \ding{183} \textbf{Introducing the security risks.} LLM-generated erroneous programs have unexpected behaviors and may crash on unknown inputs. Extensive LLM-generated erroneous programs remain a widespread and growing risk to the software infrastructure.

\begin{figure}[t]
\centering
\includegraphics[width=\linewidth]{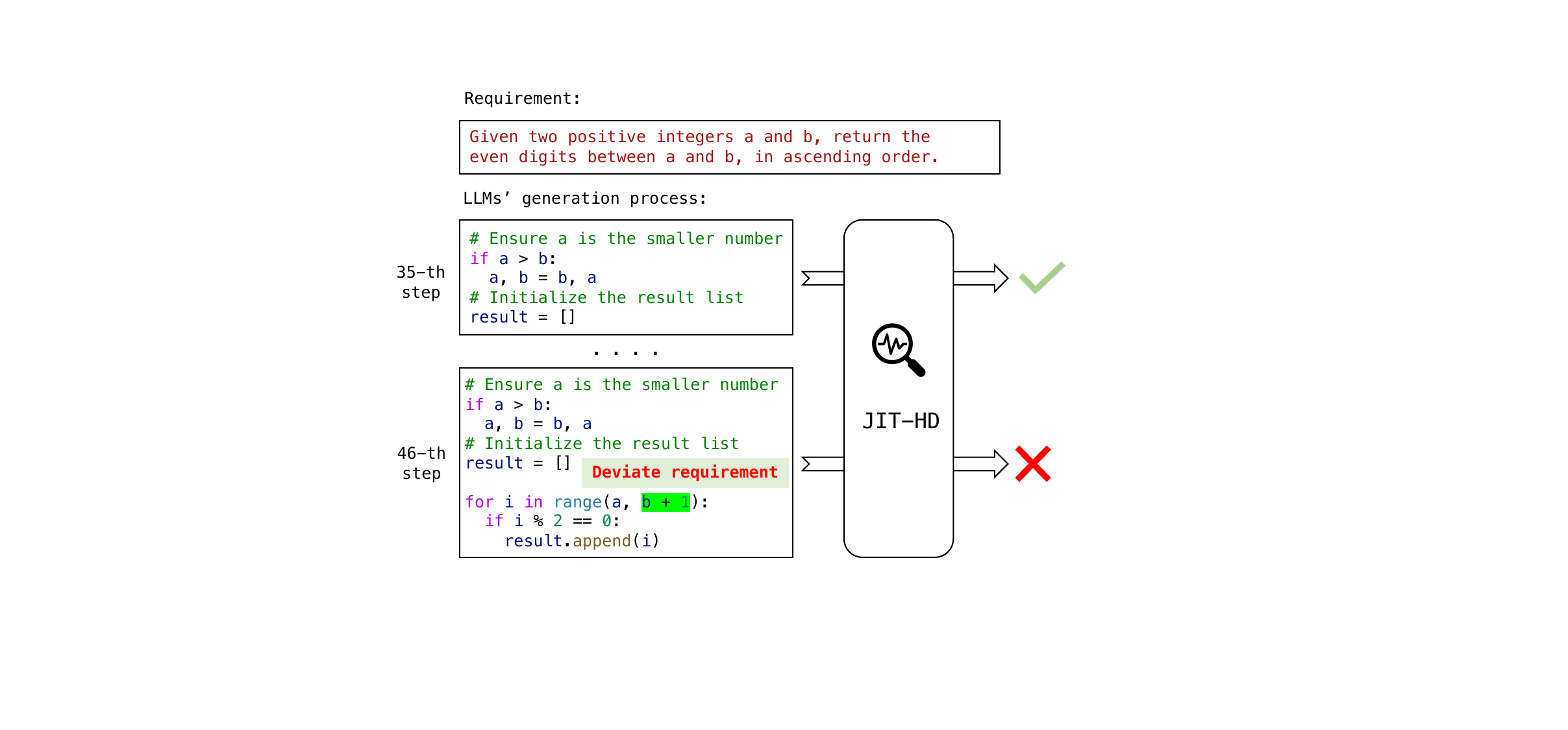}
\caption{The comparison of (a) previous LLM-based code generation approaches and (b) our \method. Previous approaches indiscriminately show developers the generated code, including the erroneous code. In contrast, \method selectively shows the generated code based on LLMs' confidence. When LLMs are uncertain, \method outputs ``\texttt{I can not solve this requirement}''.}
\label{fig:running_example}
\vspace{-0.2cm}
\end{figure}

\textbf{To address the above limitations, we propose \method, a novel LLM-based code generation approach.} Our motivation is that a critical aspect of human intelligence is to express confidence in a problem \cite{Human_Intell}. The confidence provides valuable insights into the reliability of responses. Inspired by this observation, \method aims to estimate LLMs' confidence and selectively show the LLM-generated code based on the confidence. The higher the confidence, the greater the probability that LLMs will generate the correct code for requirements.
Figure \ref{fig:running_example} compares previous code generation approaches and \method. Previous approaches \cite{DeepSeek-Coder,AceCoder,SCoT} indiscriminately show developers LLM-generated programs, including erroneous programs. In contrast, \method only shows the generated programs when LLMs are confident. When LLMs are uncertain, the generated programs probably are erroneous, and \method just outputs ``\texttt{I can not solve this requirement}''. We think that \method has the following advantages:
\begin{itemize}[leftmargin=*]
    \item \textbf{Improving developers' productivity.} \method builds more efficient human-machine collaborative programming, where LLMs only output trustworthy programs and refuse to answer uncertain requirements. It saves developers' energy spent in reading and testing LLMs' erroneous programs. Developers can focus on solving complex requirements and leave other requirements to LLMs. 
    \item \textbf{Profound implications.} \method reveals the confidence of LLMs in code generation and inspires many promising code generation techniques, \eg selective retrieval-augmented generation and human-in-the-loop generation.
    The idea of \method can also be applied to other code-related tasks (\eg vulnerability detection \cite{Devign}) and motivate broad discussions. In Section \ref{sec:conclusion}, we discuss these implications in detail.
\end{itemize} 

\textbf{The main challenge in implementing \method is how to estimate the LLMs' confidence in code generation.} The existing similar work \cite{Code_Confidence} is initially designed for natural language generation and has shown sub-optimal performance in code generation. Therefore, we propose a new confidence estimation approach. Our motivation is that high confidence typically leads to consistent answers. For example, if humans are confident about a question (\eg What shape is the Earth?), they are convinced of the answer (\eg sphere) regardless of how many times they are asked. Otherwise, if uncertain, they might arbitrarily make up disparate answers (\eg cube, pyramid). A similar phenomenon has been observed in LLMs \cite{Confidence_Fidelity,Halluc_Detect_Nature}. Inspired by this observation, our approach estimates LLMs' confidence in two steps. First, given a requirement, we independently sample multiple programs from LLMs. These programs represent LLM's multiple attempts to solve the requirement. Second, we employ a \textit{multi-modal estimator} to comprehensively measure the similarities between programs from multiple modalities (\eg lexical, syntactic, and semantic levels). The average of code similarities is considered the LLMs' confidence in the requirement.

\textbf{To evaluate \method, we collect and release an evaluation benchmark named \bench.} \bench covers two programming languages (\ie Python and Java) and consists of 2,265 samples. Each sample comprises a human-written requirement and LLM-specific labels (\ie passed or failed). The labels indicate whether the specific LLMs can generate the correct programs for requirements. We study four popular LLMs, including DeepSeek-Coder-Instruct \cite{DeepSeek-Coder} (1.3B, 6.7B) and Code Llama-Instruct \cite{CodeLlama} (13B, 7B). Detailed information about \bench is provided in Section \ref{sec:benchmark}. During the evaluation, given a requirement, we use \method to predict whether to show the generated programs. We expect that \method only shows the generation results of passed requirements and refuses failed requirements. Thus, we leverage two evaluation metrics (\ie AUROC \cite{AUROC_API} and AUCPR \cite{AUCPR_API}) to evaluate the correlation between \method's predictions and labels. The stronger the correlation, the more valuable the \method's predictions.

\textbf{We conduct a large-scale study upon \bench and yield the following observations.} \ding{182} \method can effectively estimate LLMs' confidence and accurately determine the functional correctness of generated programs. For example, \method outperforms the state-of-the-art baseline by 27.79\% in AUROC and 63.74\% in AUCPR. \ding{183} \method can decrease the number of erroneous programs shown to developers. Compared to eight baselines, it can show more correct programs and fewer erroneous programs. \ding{184} We explore the other eight variants of \method and compare them to \method. The results show the superiority \method. \ding{185} We measure the time efficiency of \method. Compared to showing code indiscriminately, \method adds slight time overhead (approximately 0.4 seconds). \ding{186} We discuss future directions to facilitate the application of LLMs in software development. We hope this work can motivate broad discussions about measuring the reliability of LLMs’ outputs in performing code-related tasks.

Our contributions are summarized below.
\begin{itemize}[leftmargin=*]
    \item We propose a novel code generation approach named \method, which selectively shows LLM-generated programs to developers based on LLMs' confidence. 
    \item To implement \method, we propose a new approach to estimating the LLMs' confidence in code generation. The confidence provides valuable insights into the correctness of the generated programs.
    \item We release a multilingual benchmark - \bench, to evaluate \method and facilitate future research. It comprises 2,265 samples and covers two programming languages and four popular LLMs.
    \item We conduct a large-scale study on \bench and show the effectiveness of \method. We discuss the future directions of our work to facilitate the application of LLMs in software development.
\end{itemize}
\vspace{-0.2cm}
\section{Motivating Examples}
\label{sec:motivating_ex}

In this section, we describe our motivations through some statistics.

\textbf{Existing LLMs may generate erroneous programs.} We select a popular LLM - DeepSeek Coder-6.7B (abbr. DS-Coder) and evaluate it on two popular benchmarks (\ie HumanEval \cite{Codex} and MBPP \cite{MBPP}). DS-Coder generates a program per requirement in benchmarks.
The evaluation results are shown in Figure \ref{fig:running_example} (a). DS-Coder fails to solve 41\% and 34\% of requirements in both benchmarks, respectively. The main reason for the failed programs is functional errors. 

\begin{figure}[t]
\centering
\includegraphics[width=\linewidth]{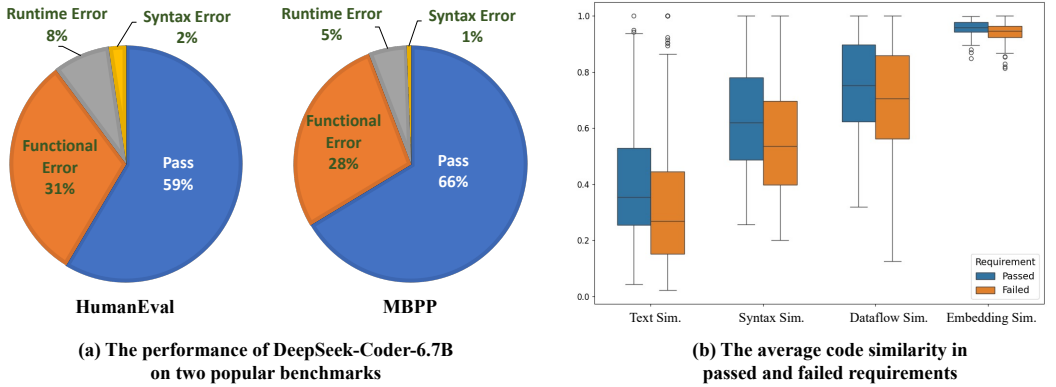}
\caption{(a) The performance of a popular LLM - DeepSeek Coder-6.7B on two popular benchmarks \cite{Codex,MBPP}. It may generate programs with errors (\eg functional errors). (b) The similarity distribution between programs in passed/failed requirements. The programs in passed requirements are more deterministic (\ie higher similarity) than the ones in failed requirements.}
\label{fig:motivating_example}
\vspace{-0.2cm}
\end{figure}

\textbf{Showing the erroneous programs will waste developers' energy.} Reading a program takes ten times longer than writing it \cite{Clean_Code}. In practice, developers spend more energy reading and testing erroneous programs than writing correct versions. 
Besides, many programs are not equipped with testing environments during development. This makes it more difficult to detect these erroneous programs.

To address the above limitation, we propose a novel code generation approach named \method. \method selectively shows LLM-generated programs based on LLMs' confidence. The confidence indicates the probability that LLMs generate correct programs. To implement \method, we propose a approach to estimating LLMs' confidence in code generation. Next, we describe the motivation for our confidence estimation approach.

\textbf{LLMs tend to generate deterministic programs when they are confident of solving the requirements.} Our idea is inspired by an observation that if humans are confident about a question (\eg In which year was the first public release of Python released?), their answers are deterministic (\eg 1991) despite how many times they are asked. Otherwise, if humans are uncertain, they might make up disparate answers (\eg 1989, 1994). LLMs also show similar behavior on question answering \cite{Confidence_Fidelity,Halluc_Detect_Nature}.

To validate that the above phenomenon also occurs in code generation, we collect the programs generated by DS-Coder in HumanEval and MBPP. DS-Coder generates $N$ ($N=20$ in this paper) programs for each requirement. The temperature for sampling is 1. If at least one of $N$ programs passes all test cases, this requirement is passed. Otherwise, it is failed. Then, we compute the average similarities between $N$ programs in passed and failed requirements as follows.
\begin{equation}
\text{AvgSim} = \frac{1}{N(N-1)} \sum_{i=1}^{N} \sum_{j=1}^{N}  \text{Sim}(c_i, c_j) \quad i \neq j
\end{equation}
where $c_i$ and $c_j$ are the $i$-th and $j$-th generated program for a requirement. $\text{Sim}(\cdot)$ is a similarity metric. Inspired by related studies \cite{ReAcc,CodeBLEU}, we set four similarity metrics: text similarity, syntax similarity, dataflow similarity, and embedding similarity. They measure the similarity between two programs from different modalities (\ie code sequences, abstract syntax trees, data flows, and embedding vectors). The detailed definitions of these metrics are in Section \ref{sec:method:confidence}.

The results are shown in Figure \ref{fig:motivating_example} (b). \textbf{We can see that the code similarities in passed requirements are greater than the ones in failed requirements.} It means that LLMs are prone to generate more deterministic (\ie more similar) programs when their outputs are correct. Based on this finding, we propose to estimate the confidence of LLMs based on the similarity between generated programs. The higher the code similarity, the greater the confidence of LLMs.
\section{\method}
\label{sec:method}

This section describes the proposed \method. We show an overview of \method and then describe its details.

\subsection{An Overview of \method}
\label{sec:method:overview}

\begin{figure*}[t]
\centering
\includegraphics[width=\linewidth]{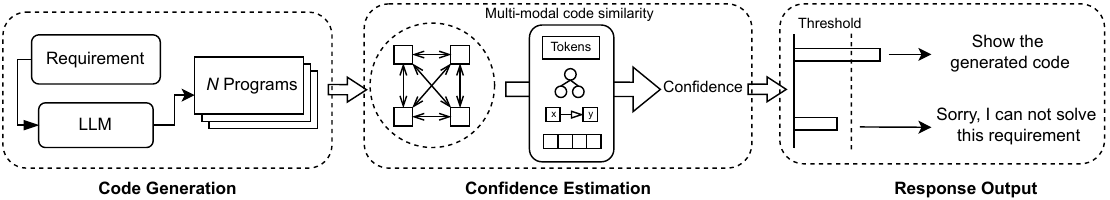}
\caption{An overview of \method. Given a requirement, it samples multiple programs from LLMs (Section \ref{sec:method:code_gen}). Then, it leverages an estimator to estimate the LLMs' confidence by measuring the multi-modal similarities between sampled programs (Section \ref{sec:method:confidence}). Finally, it determines whether to show the generated programs based on confidence (Section \ref{sec:method:response}).}
\label{fig:overview}
\vspace{-0.3cm}
\end{figure*}

\method is an LLM-based code generation approach. Unlike previous approaches, given a requirement, it can estimate the LLMs' confidence in the requirement and selectively show the generated programs to developers. Figure \ref{fig:overview} shows an overview of \method. Its pipeline consists of the following three stages:
\begin{itemize}[leftmargin=*]
    \item \textbf{Code Generation (Section \ref{sec:method:code_gen}).} Given a requirement, we leverage an LLM to generate $N$ ($N > 1$) programs. These programs denote multiple attempts by the LLM to solve the requirement.
    \item \textbf{Confidence Estimation (Section \ref{sec:method:confidence}).} A multi-modal estimator is designed to estimate the confidence of LLMs by measuring the multi-modal similarity between sampled programs. The higher the similarity, the greater the confidence of LLMs.
    \item \textbf{Response Output (Section \ref{sec:method:response}).} Based on the confidence of LLMs, it determines whether to show the generated programs. When the confidence is greater than a threshold, it shows programs. Otherwise, it outputs a refusal.
\end{itemize}

\subsection{Code Generation}
\label{sec:method:code_gen}

In this stage, we leverage an LLM to generate $N$ ($N > 1$) programs for a requirement. The generated programs denote multiple attempts by the LLM to solve the requirement. Specifically, we employ a popular sampling algorithm - temperature sampling \cite{Temperature_Sampling}. The motivations for choosing temperature sampling are two-fold. First, temperature sampling introduces randomness into the decoding process and can sample diverse answers based on LLMs' logits. It is beneficial to reveal the uncertainty of LLMs \cite{Calibrating_Fidelity,Halluc_Detect_Nature}. Second, temperature sampling is a mainstream way to liberate the coding ability of LLMs and has been widely employed in code generation \cite{CodeLlama,DeepSeek-Coder,DeepSeek-Coder-V2}.

We want to clarify that the above sampling aims to facilitate confidence estimation and is invisible to developers. If \method finally determines that LLMs are uncertain, these sampled programs will not be shown to developers. Besides, sampling can be parallelized, so it does not add extra time overheads. We measure the time efficiency of \method in Section \ref{sec:discussion:efficiency} and tune the default sampling size $N$ to 20.

\subsection{Confidence Estimation}
\label{sec:method:confidence}

\begin{algorithm}[t]
\centering
\caption{The workflow of our estimator.}
\label{alg:estimator}
\begin{algorithmic}[1]
\STATE \textbf{Input:} A set of sampled programs $\mathbf{C}$
\STATE \textbf{Output:} A confidence score
\STATE $N \gets \operatorname{Len}(\mathbf{C})$ \textit{// $\operatorname{Len}()$ is a function returning the number of elements in a list}
\STATE $Sim\_list \gets []$
\FOR{$i = 1$ to $N$}
    \FOR{$j = 1$ to $N$}
        \IF{$i \neq j$}
            \STATE $c_i \gets \mathbf{C}[i]$, $c_j \gets \mathbf{C}[j]$
            \STATE Compute $\text{Sim}(c_i, c_j)$ Using Equation \ref{equ:hybrid_sim}
            \STATE Append $\text{Sim}(c_i, c_j)$ to $Sim\_list$
        \ENDIF
    \ENDFOR
\ENDFOR
\STATE $Confidence \gets \operatorname{SUM}(Sim\_list) / \operatorname{Len}(Sim\_list)$
\RETURN $Confidence$
\end{algorithmic}
\end{algorithm}

This stage aims to estimate the LLMs' confidence in code generation. The confidence reflects the probability that LLMs can generate the correct programs for the requirement. As discussed in Section \ref{sec:motivating_ex}, LLMs tend to generate deterministic programs when they are confident. Conversely, the generated programs typically are disparate when LLMs are uncertain. Thus, our estimator estimates LLMs' confidence by measuring the similarity between sampled programs.

The key challenge in implementing the estimator is to accurately measure the similarity between programs. We think that a good estimator should satisfy two principles:
\begin{itemize}[leftmargin=*]
    \item \textbf{Comprehensive.} The similarity between programs is often reflected at different levels, \eg text surfaces, syntactic structures and semantic embeddings. Thus, our estimator should measure the code similarity in a comprehensive view. A narrow perspective may result in false results. 
    \item \textbf{General.} LLMs are emerging and employed to generate programs across various domains. Therefore, our estimator should be generalizable across domains and easily applicable to different LLMs.
\end{itemize}

However, no existing studies fulfill both principles. Traditional code similarity metrics (\eg BLEU \cite{BLEU} and CodeBLEU \cite{CodeBLEU}) fall into narrow perspectives (\eg text surfaces or syntactic structures) and struggle to provide comprehensive measurements. Deep learning-based approaches are typically task-specific (\eg code clone detection \cite{ZC3}) and lack generalization.

To satisfy the above principles, we propose a multi-modal estimator. It comprehensively measures the code similarity from multiple modalities and can generalize to various LLMs. Algorithm \ref{alg:estimator} shows the workflow of our estimator. Specifically, it parses the code into four mainstream modalities: token sequences, concrete syntax trees, data flows, and semantic embeddings. Then, it computes a weighted sum of similarities in different modalities:
\begin{equation}
\label{equ:hybrid_sim}
\begin{aligned}
        \text{Sim}(c_i, c_j) = & \alpha \times \text{Sim}_{\text{text}} (c_i, c_j) + \beta \times \text{Sim}_{\text{syntax}}(c_i, c_j) + \\
        & \gamma \times \text{Sim}_{\text{df}} (c_i, c_j) + \delta \times \text{Sim}_{\text{embed}} (c_i, c_j)
\end{aligned}
\end{equation}
where $c_i$ and $c_j$ are two generated programs. $\text{Sim}_{\text{text}}$, $\text{Sim}_{\text{syntax}}$, $\text{Sim}_{\text{df}}$, and $\text{Sim}_{\text{embed}}$ denote the similarities in four modalities, respectively. 
$\alpha$, $\beta$, $\gamma$, and $\delta$ are hyper-parameters and within the range $[0,1]$. We tune these hyper-parameters on the training data and release them in our replication package \cite{Anonymity2024}.
The LLMs' confidence in a requirement is computed as the average of the pairwise similarity scores between the generated programs. We describe how to compute similarities in different modalities as follows.

\textbf{Text Similarity.} Programs with similar functionalities often have similar text surfaces. Researchers have proposed many successful metrics to measure the text similarity between programs \cite{BLEU, CrystalBLEU}. Following existing studies \cite{BLEU}, we split programs into token sequences and compute the $n$-gram similarity between two code sequences. Specifically,
\begin{equation}
\text{Sim}_{\text{text}}(c_i, c_j) = \exp \left(\sum_{n=1}^{4} \frac{1}{4} \log \frac{s^n_{ij}}{s^n_j}\right)
\end{equation}
where the maximum of $n$ is empirically set to 4. $s^n_{ij}$ is the number of overlapping $n$-grams between $c_i$ and $c_j$. $s^n_j$ means the number of $n$-grams in $c_j$.

\textbf{Syntactic Similarity.} The code is structured and can be parsed into tree structures, \eg Concrete Syntax Trees (CSTs). Programs with different text surfaces may have similar syntactic structures. Following the related work \cite{CodeBLEU}, we calculate the similarity in syntactic structures. Specifically, we parse the programs into CSTs and extract sub-trees from CSTs. Then, we compute the ratio of overlapping sub-trees between two programs. Formally,
\begin{equation}
\text{Sim}_{\text{syntax}}(c_i, c_j) = \frac{|\text{CST}(c_i) \cap \text{CST}(c_j)|}{|\text{CST}(c_j)|}
\end{equation}
where $\text{CST}(\cdot)$ is a function for extracting all CST sub-trees of a program. 

\textbf{Data Flow Similarity.} The code is executable and contains rich data flow information. Data flows reflect the dependencies between variables and can represent the code's functionality \cite{GraphCodeBERT}. Inspired by related works \cite{GraphCodeBERT,CodeBLEU}, we first extract the data-flow graphs from programs. Each node (\eg $v_i, v_j$) in the data-flow graphs denotes a variable. The directed edge $e=\langle v_i, v_j \rangle$ from $v_i$ to $v_j$ refers that the value of $v_j$ comes from $v_i$. Then, we compute the ratio of overlapping data-flow edges between two programs. Specifically,
\begin{equation}
\text{Sim}_{\text{df}}(c_i, c_j) = \frac{|\text{DFG}_{\text{edge}}(c_i) \cap \text{DFG}_{\text{edge}}(c_j)|}{|\text{DFG}_{\text{edge}}(c_j)|}
\end{equation}
where $\text{DFG}(\cdot)$ is a function for extracting all data flow edges of a program. 

\textbf{Semantic Embedding Similarity.} LLMs have shown impressive abilities in code representation \cite{Code_Understanding_LLM}. Thus, we can leverage LLMs to represent a program with a semantic embedding (\ie a high-dimension vector). Then, we compute the cosine similarity between embeddings as follows:
\begin{equation}
\label{equ:embedding_sim}
\begin{gathered}
\mathbf{e}_i = \text{M}(c_i), \mathbf{e}_j = \text{M}(c_j) \\
\text{Sim}_{\text{embed}}(c_i, c_j) = \frac{\mathbf{e}_i\cdot\mathbf{e}_j}{\|\mathbf{e}_i\| \|\mathbf{e}_j\|}
\end{gathered}
\end{equation}
$\text{M}$ denotes an LLM used for encoding programs. $\mathbf{e}_i$ and $\mathbf{e}_j$ denote the semantic embeddings of $c_i$ and $c_j$, respectively. 

Specifically, we leverage a powerful LLM (\eg DeepSeek Coder-6.7B-base \cite{DeepSeek-Coder}) as $\text{M}$ and pick the last token's hidden state in the last layer as the embedding. Our motivations are two-fold. First, because LLMs are autoregressive, only the last code token's hidden state is computed based on the whole code. Thus, it can represent the comprehensive semantics of the code. Second, the shallow layers of LLMs mainly capture the lexical information of programs, while the deep layer focuses on the semantic information of programs \cite{Transformer_Analysis}. Thus, we select the hidden states in the last layer.

Our estimator provides a \textit{comprehensive} measurement of code similarity and is \textit{general} for different LLMs and programs. In Section \ref{sec:results:RQ3_estimator}, we compare our estimator to eight other variants (\eg BLEU-based, CodeBLEU-based) and validate the superiority of our estimator.

\subsection{Response Output}
\label{sec:method:response}

In this stage, we determine whether to show the generated programs to developers based on LLMs' confidence. Specifically, if the LLMs' confidence is greater than a threshold $T$, \method shows the sampled programs in the first stage to developers. Otherwise, \method outputs a refusal. For example, \method will output a special response - \texttt{Sorry, I cannot solve this requirement.} In the future, researchers can design more advanced generation techniques when LLMs are uncertain, \eg human-in-the-loop generation and retrieval-augmented generation.
Because this paper is an early step in discussing selective code generation, we leave this exploration in future work.

In practice, practitioners can control the aggressiveness of \method by adjusting the threshold $T$. A larger $T$ makes \method more conservative, and LLMs only show the generation results of requirements with very high confidence. In contrast, a smaller $T$ makes \method more aggressive, and the shown programs may contain minor mistakes.

\section{Evaluation Benchmark - \bench}
\label{sec:benchmark}

In this section, we collect an evaluation benchmark named \bench to evaluate \method. We show an overview of \bench and then describe its collection pipeline.

\subsection{An Overview of \bench}
\label{sec:benchmark:overview}

\begin{figure}[t]
\centering
\includegraphics[width=0.9\linewidth]{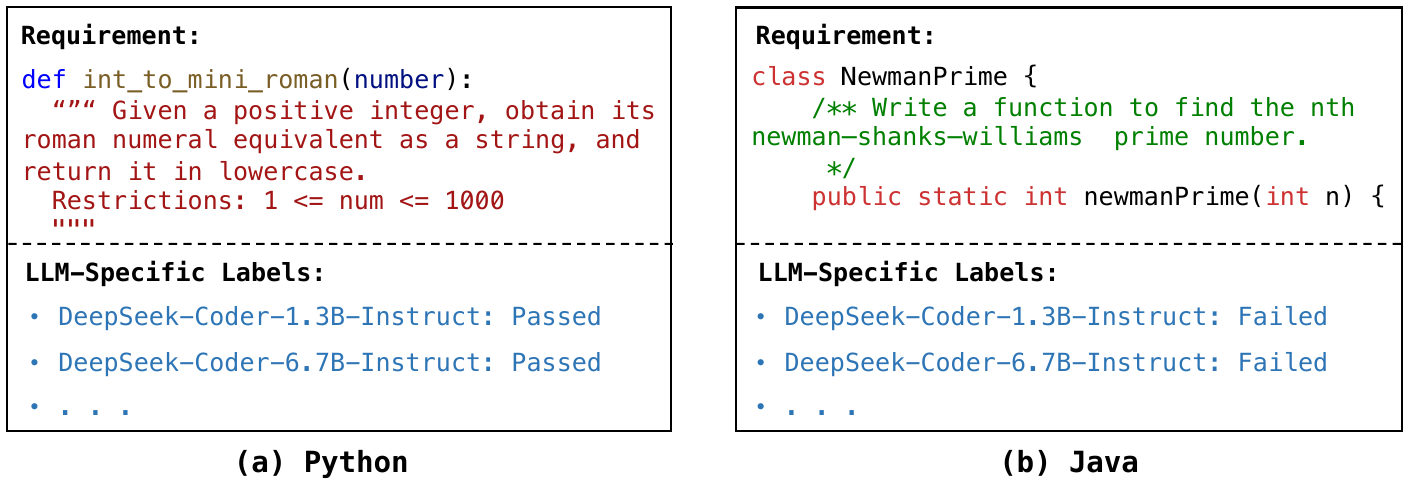}
\vspace{-0.2cm}
\caption{Two samples of \bench in Python and Java, respectively. Each sample has two components: a requirement and LLM-specific binary labels (\ie passed or failed). The label represents whether a specific LLM can generate the correct programs for a requirement. We expect that \method only shows the generated code for passed requirements and refuses failed requirements.}
\label{fig:benchmark_example}
\vspace{-0.2cm}
\end{figure}

\textbf{Data Format.} \bench covers two popular programming languages, \ie Python and Java. Both languages are mainstream programming languages for software developers and are widely used by developers around the world \cite{Python_Java}. Thus, we start \bench with these two languages and leave other languages in future work.
Figure \ref{fig:benchmark_example} shows two samples in \bench. Each sample in \bench consists of two components: a requirement and LLM-specific binary labels (\ie passed or failed). The label represents whether the specific LLM can generate the correct programs for a requirement. We expect that \method only shows the generated code for passed requirements and refuses failed requirements.

\textbf{Data Statistics.} The statistics of \bench are shown in Table \ref{tab:benchmark_statistic}. For each language, \bench contains a training set and a test set. For Python, the training set and test set both contain 569 samples. The training and test sets for Java comprise 563 and 564 samples, respectively. In this paper, we study four popular LLMs and annotate each sample with four LLM-specific labels. The details of four LLMs are described in Section \ref{sec:study_design:LLMs}. Given that the coding capabilities of LLMs vary, the label distributions of different LLMs are different.


\begin{table}[t]
\caption{The statistics of \bench. ``Train / Test'' means the number of samples in the training and test sets, respectively. ``Train/Test Distribution'' denotes the rates of passed and failed requirements in the training and test sets, respectively. DS-Coder: DeepSeek Coder. Ins: Instruct.}
\vspace{-0.2cm}
\label{tab:benchmark_statistic}
\resizebox{0.8\linewidth}{!}{
\begin{tabular}{cclcc}
\toprule
Language & Train / Test & \multicolumn{1}{c}{LLM} & Train Distribution & Test Distribution \\ \midrule
\multirow{4}{*}{Python} & \multirow{4}{*}{569 / 569} & DS-Code-6.7B-Ins & 41.12\% / 58.88\% & 40.25\% / 59.75\% \\
 &  & DS-Code-1.3B-Ins & 35.33\% / 64.67\% & 30.23\% / 69.77\% \\
 &  & Codellama-13B-Ins & 34.97\% / 65.03\% & 30.58\% / 69.42\% \\
 &  & Codellama-7B-Ins & 21.61\% / 78.38\% & 21.97\% / 78.03\% \\ \midrule
\multirow{4}{*}{Java} & \multirow{4}{*}{563 / 564} & DS-Code-6.7B-Ins & 39.96\% / 60.04\% & 51.95\% / 48.05\% \\
 &  & DS-Code-1.3B-Ins & 40.67\% / 59.33\% & 43.62\% / 56.38\% \\
 &  & Codellama-13B-Ins & 28.24\% / 71.76\% & 33.33\% / 66.67\% \\
 &  & Codellama-7B-Ins & 20.60\% / 79.40\% & 25.35\% / 74.65\% \\
 \bottomrule
\end{tabular}}
\vspace{-0.2cm}
\end{table}

\subsection{Benchmark Collection Pipeline}
\label{sec:benchmark:collection}

The collection pipeline of \bench consists of the following three stages.

\textbf{Step \ding{182}: Requirement Collection.} We collect requirements from two popular benchmarks, \ie HumanEval-X \cite{Codex} and MBXP \cite{MBPP}. These requirements are written by human developers and cover diverse programming domains (\eg data science and text processing). We collect requirements in Python and Java and leave other languages in future work. We randomly split the requirements into training and test sets in a 5:5 ratio. The split ratio is empirically set based on related work \cite{APPS}.

\textbf{Step \ding{183}: Label Annotation.} For each requirement, we annotate LLM-specific labels. Specifically, we allow an LLM to generate $K$ programs for a requirement and execute test cases to check the programs. If at least one of the $K$ programs passes all test cases, the requirement is passed; otherwise, it is failed.

In practice, the performance of LLMs on code generation is impacted by inference settings, \eg sampling algorithms \cite{AdapT} and prompt designs \cite{AceCoder,SCoT}. Different inference settings often lead to different generation results for the same requirement and LLM. To ensure annotation consistency, we interviewed 10 experienced developers who regularly use LLMs and summarized their conventions. For example, developers often employ zero-shot prompting and sample programs from LLMs using temperature sampling. The common temperatures include 0, 0.2, 0.6, 0.8, and 1. Besides, developers' time and energy are limited. If an LLM fails to solve a requirement within a limited number of attempts (\ie average: 3.4, max: 5), developers prefer to manually write the code than read the generated erroneous programs.

Based on the above conventions, we design a default inference setting. Given a requirement, we use zero-shot prompting and temperature sampling to sample five programs from an LLM. The five programs are generated by different temperatures (\ie 0, 0.2, 0.6, 0.8, and 1). If any of the five programs pass all test cases, the requirement is passed for specific LLMs; otherwise, it is failed. In future work, practitioners can design inference settings based on their application scenarios and label data following our pipeline.
\section{Study Design}
\label{sec:study_design}

We conduct a large-scale study to evaluate our \method. This section describes the settings of our study, including research questions, studied LLMs, compared baselines, and evaluation metrics.

\subsection{Research Questions}
\label{sec:study_design:RQs}

The goal of \method is to estimate LLMs' confidence in requirements and avoid showing the generated erroneous code to developers. Thus, we design the RQ1 and RQ2 to evaluate the effectiveness of \method in these two aspects.

\textbf{RQ1: How does \method perform in distinguishing between passed and failed requirements?}

As shown in Figure \ref{fig:overview}, \method estimates LLMs' confidence and only shows the generated programs for high-confidence requirements. Thus, accurate confidence estimation is crucial. In this RQ, we evaluate the capability of \method in distinguishing between passed and failed requirements based on confidence. 

Specifically, we apply \method to \bench and estimate a confidence score for each requirement. If the confidence score exceeds a threshold, the requirement is predicted as passed. Otherwise, it is predicted as failed. Then, we use evaluation metrics (Section \ref{sec:study_design:metrics}) to comprehensively measure the performance of \method under different thresholds. To show the superiority of \method, we design eight baselines (Section \ref{sec:study_design:baselines}) for comparison.

\textbf{RQ2: How does \method perform in decreasing the number of erroneous programs shown to developers?}

In this RQ, we set three control groups. In the first group, LLMs indiscriminately show the generated programs to developers. In the second and third groups, LLMs selectively show the generated programs based on the predictions of baselines and \method, respectively. Then, we count the numbers of correct and erroneous programs shown to developers in three groups. 

As stated in Section \ref{sec:method:confidence}, a critical stage in \method is to estimate the LLMs' confidence by measuring the similarity between the generated programs. We propose to measure the similarity from multiple modalities. The RQ3 is designed to validate the effectiveness of our estimator and explore the contributions of different modalities.

\textbf{RQ3: How effective is the multi-modal estimator?}

We design eight variants of our estimators, which use different code similarity metrics. Then, we compare them to our estimator on four LLMs and analyze the contributions of different modalities in our estimator.

\vspace{-0.2cm}
\subsection{Studied LLMs}
\label{sec:study_design:LLMs}

This paper studies four popular LLMs, including DeepSeek-Coder-Instruct-\{6.7B, 1.3B\}, and Code Llama-Instruct-\{13B, 7B\}. Their details are shown as follows.

\textbf{DeepSeek-Coder-Instruct (DS-Coder-Ins) \cite{DeepSeek-Coder}.} It is a family of LLMs for code released by DeepSeek-AI. Researchers develop DS-Coder-Ins by enhancing the DS-Coder-Base through instruction tuning on high-quality data. The training data comprises helpful and impartial human instructions structured by the single-turn dialogue format, containing 2B tokens in total. DS-Coder-Ins provides multiple versions with different sizes. In our experiments, we select DS-Code-Ins with 6.7B and 1.3B parameters.

\textbf{Code Llama-Instruct (CodeLlama-Ins) \cite{CodeLlama}.} It is a popular family of LLMs for code developed by Meta-AI. The first part of its training data is a proprietary dataset consisting of many multi-turn dialogues between a user and an assistant. The second part of the training data is 14k code-related question-answering pairs generated by Llama 2-70B \cite{Llama-2}. In this paper, we select CodeLlama-Ins with 7B and 13B parameters.

\subsection{Compared Baselines}
\label{sec:study_design:baselines}

To our knowledge, this paper is the first to explore selectively showing LLM-generated programs to developers. We collect some studies with motivations similar to this paper and apply them to our task. These related studies can be divided into two groups, including code confidence estimation and requirement confidence estimation. We describe their details as follows.

\textbf{Code confidence-based approaches} aim to estimate the LLMs' confidence in the generated programs and average the confidence of all programs as output. If the outputted confidence is greater than a threshold, the programs are shown to developers. A recent work \cite{Code_Confidence} studied a few approaches to estimating code confidence. We select four representative approaches as baselines. We describe their details as follows.
\begin{itemize}[leftmargin=*]
    \item \textbf{Average Prob.} It collects the generation probabilities of code tokens and averages all probabilities as the confidence.  
    \item \textbf{Product Prob.} It collects the generation probabilities of code tokens and calculates the product of all probabilities as the confidence.
    \item \textbf{Self-Asking (Code).} It directly uses prompting to ask for the confidence of LLMs in the code. Specifically, we craft a prompt asking LLMs to determine the functional correctness of the generated code, \ie answering ``Yes'' or ``No''. The prompt is available in our replication package \cite{Anonymity2024}. Then, we extract the output probability of ``Yes'' as the confidence.
    \item \textbf{Code Classifier.} It is a neural classifier that takes the generated code as input and predicts a probability that the code is functionally correct. Specifically, we select a popular code representation model (\ie UniXcoder \cite{UniXcoder}) as the backbone and add a classification head to it. We split the original training set into a new training and validation set. We train this classifier for 20 epochs and select the best checkpoint based on the performance upon the validation set. 
\end{itemize}

\textbf{Requirement confidence-based approaches} are to estimate the LLMs' confidence in requirements and determine whether to generate programs for requirements. We apply three effective approaches \cite{Self-Guided-RAG} in question answering to code generation and design the following baselines.
\begin{itemize}[leftmargin=*]
    \item \textbf{Self-Asking (Re.).} It directly uses prompting to ask for the LLMs' confidence in successfully solving current requirements. Specifically, we craft a prompt asking LLMs to determine whether they can solve a requirement, \ie answering ``Yes'' or ``No''. The prompt is available in our replication package \cite{Anonymity2024}. Then, we extract the generation probability of ``Yes'' as the confidence.
    \item \textbf{K-Nearest Neighbor Search (K-NNS).} Its motivation is that if LLMs can solve a specific requirement, then a similar requirement is likely solvable. Given a requirement, K-NNS retrieves $K$ most similar requirements in the training set. Then, it calculates the ratio of passed requirements in retrieved results as output. This paper employs two retrieval metrics, \ie BM25 \cite{BM25} and embedding similarity. BM25 is a common text similarity metric. We use a natural language representation model (\ie Sentence-BERT \cite{Sentence-BERT}) to convert requirements into embedding vectors and compute the cosine similarity between two embeddings. The value of $K$ is tuned on the performance of K-NNS on the training set.
    \item \textbf{Requirement Classifier.} It is a neural classifier to predict the probability that LLMs successfully solve the requirement. We select a natural language representation model (\eg Sentence-BERT \cite{Sentence-BERT}) as the backbone and add a classification head to it. We split the original training set into a new training and validation set. We train this classifier for 20 epochs and select the best checkpoint based on the performance upon the validation set. 
\end{itemize}

\subsection{Evaluation Metrics}
\label{sec:study_design:metrics}

The problem studied in this paper can be viewed as a binary classification task, \ie whether to show LLM-generated programs to developers. Traditional metrics for classification tasks (\eg accuracy and F1) are sensitive to pre-defined thresholds and cannot reflect the comprehensive performance of approaches. Following previous studies \cite{Confidence_Study,Halluc_Detect_Nature}, we employ two popular metrics, \ie AUROC and AUCPR. They can measure the comprehensive performance of approaches under different thresholds. We describe the details of two metrics as follows.

\textbf{AUROC} is a popular metric for evaluating binary classifiers. The closer the AUROC is to 1, the stronger the classification ability of a classifier. We compute the AUROC through a public API \cite{AUROC_API}.

AUROC is the area under a ROC curve. The ROC curve represents the relationships between TPR (True Positive Rate) and FPR (False Position Rate) under different thresholds. The calculation formulas for TPR and FPR are as follows:
\begin{equation}
\text{TPR} = \frac{\text{TP}}{\text{TP}+\text{FN}}, \quad
 \text{FPR} = \frac{\text{FP}}{\text{FP}+\text{TN}}
\end{equation}
where TP, FP, TN, and FN denote the number of true positives, false positives, true negatives, and false negatives, respectively. 

\textbf{AUCPR.} It is an effective classification task metric and particularly suitable for unbalanced datasets. AUCPR is a value between 0 and 1. The larger the AUCPR, the better the classification ability of the model. We calculate the AUCPR using a public API \cite{AUCPR_API}.

Specifically, AUCPR is the area under a PR (Precision-Recall) Curve. The PR curve reflects the relationships between precision and recall under different thresholds. The calculation formulas for precision and recall are as follows:
\begin{equation}
\text{Precision} = \frac{\text{TP}}{\text{TP}+\text{FP}}, \quad
 \text{Recall} = \frac{\text{TP}}{\text{TP}+\text{FN}}
\end{equation}

\section{Results and Analyses}
\label{sec:results}

\subsection{RQ1: The performance in distinguishing between passed and failed requirements}
\label{sec:results:RQ1}

As shown in Figure \ref{fig:overview}, \method estimates LLMs' confidence and only shows the generated programs for high-confidence requirements. Thus, accurate confidence estimation is crucial. In the RQ1, we evaluate the effectiveness of the confidence scores from \method in distinguishing between passed and failed requirements. 

\begin{table}[t]
\centering
\caption{The performance of different approaches in distinguishing between passed and failed requirements on \bench (Python).}
\vspace{-0.2cm}
\label{tab:RQ1_python}
\resizebox{\linewidth}{!}{
\begin{tabular}{lcccccccc}
\toprule
\multicolumn{1}{c}{} & \multicolumn{2}{c}{CodeLlama-13B-Ins} & \multicolumn{2}{c}{CodeLlama-7B-Ins} & \multicolumn{2}{c}{DS-Coder-6.7B-Ins} & \multicolumn{2}{c}{DS-Coder-1.3B-Ins} \\
\multicolumn{1}{c}{\multirow{-2}{*}{Approach}} & AUROC & AUCPR & AUROC & AUCPR & AUROC & AUCPR & AUROC & AUCPR \\ \midrule
Average Prob. & 56.64\% & 36.30\% & 58.36\% & 27.27\% & 54.75\% & 44.96\% & 53.49\% & 34.90\% \\
Product Prob. & 54.33\% & 34.71\% & 57.12\% & 25.76\% & 52.11\% & 42.85\% & 51.33\% & 33.45\% \\
Self-Asking (Code) & 53.15\% & 33.11\% & 56.17\% & 24.33\% & 50.97\% & 41.58\% & 50.28\% & 31.29\% \\
Code Classifier & 55.13\% & 35.19\% & 57.32\% & 25.99\% & 53.05\% & 43.77\% & 52.15\% & 33.67\% \\
\midrule
Self-Asking (Re.) & 51.40\% & 32.40\% & 52.47\% & 26.07\% & 49.92\% & 41.11\% & 54.30\% & 34.78\% \\
K-NNS (BM25) & 60.14\% & 36.42\% & 57.60\% & 29.58\% & 61.48\% & 48.09\% & 62.89\% & 38.56\% \\
K-NNS (Embedding) & 60.33\% & 37.79\% & 56.89\% & 27.74\% & 59.64\% & 46.09\% & 61.90\% & 37.87\% \\
Requirement Classifier & 47.95\% & 44.37\% & 44.37\% & 18.80\% & 61.56\% & 48.83\% & 47.95\% & 28.17\% \\
\rowcolor[rgb]{ .741,  .843,  .933}
\method & \textbf{74.05\%} & \textbf{54.83\%} & \textbf{73.91\%} & \textbf{46.55\%} & \textbf{63.73\%} & \textbf{54.38\%} & \textbf{73.36\%} & \textbf{50.96\%} \\ \midrule
Relative Improvement & {\color[HTML]{F54A45} \textbf{22.74\%}} & {\color[HTML]{F54A45} \textbf{23.57\%}} & {\color[HTML]{F54A45} \textbf{26.64\%}} & {\color[HTML]{F54A45} \textbf{57.37\%}} & {\color[HTML]{F54A45} \textbf{3.53\%}} & {\color[HTML]{F54A45} \textbf{11.37\%}} & {\color[HTML]{F54A45} \textbf{16.65\%}} & {\color[HTML]{F54A45} \textbf{32.16\%}} \\
\bottomrule
\end{tabular}}
\vspace{-0.2cm}
\end{table}

\begin{table}[t]
\centering
\caption{The performance of different approaches in distinguishing between passed and failed requirements on \bench (Java).}
\vspace{-0.2cm}
\label{tab:RQ1_java}
\resizebox{\linewidth}{!}{
\begin{tabular}{lcccccccc}
\toprule
\multicolumn{1}{c}{} & \multicolumn{2}{c}{CodeLlama-13B-Ins} & \multicolumn{2}{c}{CodeLlama-7B-Ins} & \multicolumn{2}{c}{DS-Coder-6.7B-Ins} & \multicolumn{2}{c}{DS-Coder-1.3B-Ins} \\
\multicolumn{1}{c}{\multirow{-2}{*}{Approach}} & AUROC & AUCPR & AUROC & AUCPR & AUROC & AUCPR & AUROC & AUCPR \\ \midrule
Average Prob. & 55.64\% & 36.07\% & 53.38\% & 25.93\% & 51.96\% & 51.53\% & 58.90\% & 48.54\% \\
Product Prob. & 53.12\% & 34.79\% & 51.98\% & 23.78\% & 49.22\% & 48.79\% & 56.19\% & 46.78\% \\
Self-Asking (Code) & 52.19\% & 33.14\% & 50.77\% & 23.64\% & 48.56\% & 47.98\% & 55.41\% & 45.70\% \\
Code Classifier & 54.33\% & 35.67\% & 52.16\% & 24.77\% & 50.15\% & 50.77\% & 57.21\% & 47.64\% \\
\midrule
Self-Asking (Re.) & 58.22\% & 38.65\% & 56.13\% & 28.50\% & 56.52\% & 58.45\% & 62.67\% & 51.75\% \\
K-NNS (BM25) & 49.80\% & 33.58\% & 54.59\% & 28.54\% & 52.71\% & 53.87\% & 52.01\% & 44.16\% \\
K-NNS (Embedding) & 54.71\% & 36.46\% & 54.53\% & 27.24\% & 52.36\% & 56.07\% & 54.87\% & 50.00\% \\
Requirement Classifier & 53.61\% & 34.71\% & 48.78\% & 23.79\% & 41.98\% & 45.54\% & 55.46\% & 46.58\% \\ 
\rowcolor[rgb]{ .741,  .843,  .933}
\method & \textbf{68.46\%} & \textbf{53.95\%} & \textbf{71.73\%} & \textbf{46.73\%} & \textbf{64.71\%} & \textbf{64.17\%} & \textbf{68.21\%} & \textbf{59.92\%} \\ \midrule
Relative Improvement & {\color[HTML]{F54A45} \textbf{17.59\%}} & {\color[HTML]{F54A45} \textbf{39.59\%}} & {\color[HTML]{F54A45} \textbf{27.79\%}} & {\color[HTML]{F54A45} \textbf{63.74\%}} & {\color[HTML]{F54A45} \textbf{14.49\%}} & {\color[HTML]{F54A45} \textbf{9.79\%}} & {\color[HTML]{F54A45} \textbf{8.84\%}} & {\color[HTML]{F54A45} \textbf{15.79\%}} \\
\bottomrule
\end{tabular}}
\vspace{-0.2cm}
\end{table}


\textbf{Settings.} We leverage \method or baselines to compute a confidence score for each requirement in \bench. If the confidence score exceeds a threshold, the requirement is predicted as passed. Otherwise, it is predicted as failed. Then, we use evaluation metrics (Section \ref{sec:study_design:metrics}) to comprehensively measure the classification performance of different approaches under different thresholds.

\textbf{Results and Analyses.} The results of RQ1 are shown in Table \ref{tab:RQ1_python} and \ref{tab:RQ1_java}. From the results, we obtain the following insights. 

\textbf{\method substantially outperforms baselines in AUROC and AUCPR.} In terms of CodeLlama-7B-Instruct, \method outperforms the strongest baseline by up to 27.79\% in AUROC and 63.74\% in AUCPR. Both metrics show the comprehensive performance of approaches under different classification thresholds. The substantial improvements show that the confidence from \method provides reliable insights into the correctness of LLM-generated programs.  

\textbf{\method is effective in different LLMs and programming languages.} As shown in Table \ref{tab:RQ1_python} and \ref{tab:RQ1_java}, \method achieves stable improvements in LLMs with different scales (\eg 13B, 7B, 6.7B, and 1.3B) and languages (\ie Python and Java). In the future, practitioners can apply \method to more advanced LLMs and new languages in a plug-and-play manner. 

\textbf{Self-Asking and probability-based approaches are prone to be over-confident.} As shown in Table \ref{tab:RQ1_python} and \ref{tab:RQ1_java}, the performance of Self-Asking and probability-based approaches is not satisfactory. We inspect 50 failure samples of these baselines. We find that LLMs are prone to express high confidence in requirements and the generated programs. Similar phenomena have also been found in work in other fields \cite{Calibrating_Fidelity,Confidence_Study}. In contrast, our \method offers a more promising direction to confidence estimation and significantly outperforms baselines.

\greyboxb{Answer to RQ1: } {In four LLMs and two programming languages, \method substantially outperforms baselines in AUROC and AUCPR. The results comprehensively show the effectiveness of \method in distinguishing between passed and failed requirements.}

\begin{figure*}[t]
\centering
\includegraphics[width=\linewidth]{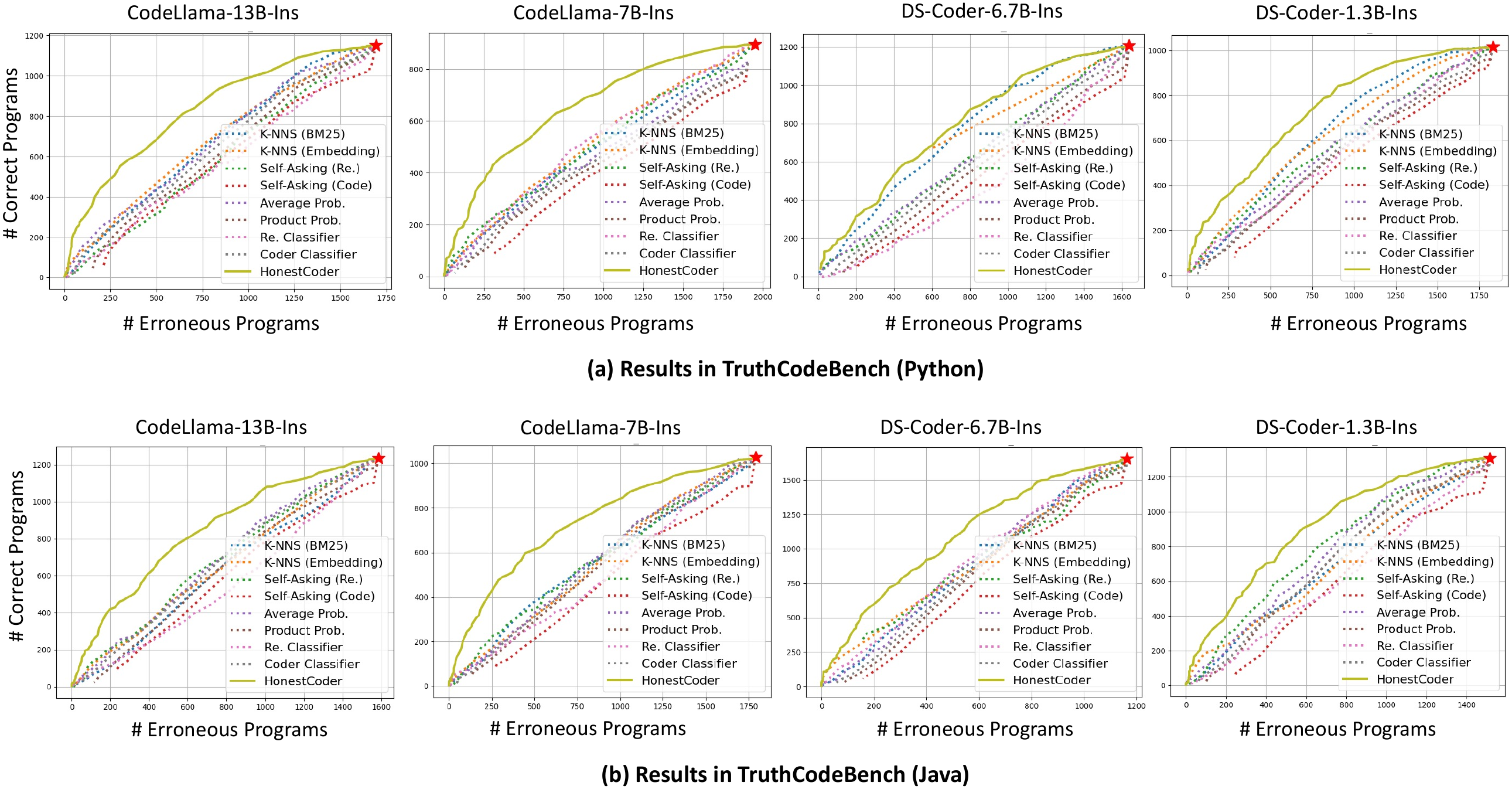}
\vspace{-0.4cm}
\caption{The number of correct and erroneous programs shown to developers on \bench. The stars denote the results of showing LLM-generated programs indiscriminately. The curves denote the results of baselines and \method under different thresholds. The closer the curve is to the upper left corner, the better the approach's performance.}
\label{fig:RQ2_results}
\end{figure*}

\subsection{RQ2: The performance in reducing erroneous programs shown to developers}
\label{sec:results:RQ2_reducing}

As stated in Section \ref{sec:intro}, the goal of \method is to avoid showing LLM-generated erroneous programs to developers. Meanwhile, we expect that \method does not hinder LLMs from showing correct programs. Thus, in the RQ2, we evaluate the impact of \method on the number of correct and erroneous programs shown to developers.

\textbf{Settings.} We set three control groups. In the first group, LLMs indiscriminately show the generated programs to developers. In the second and third groups, LLMs selectively show the generated programs based on the predictions of baselines and our \method, respectively. Then, we count the correct and erroneous programs shown to developers in three groups. According to the developers' conventions summarized in Section \ref{sec:benchmark:collection}, we ask LLMs to generate five programs per requirement. We execute test cases to check the correctness of generated programs.

\method and baselines all rely on thresholds for making predictions. To achieve a comprehensive comparison, we take 100 equally spaced values from each approach's range as its thresholds and show the results under these thresholds.

\textbf{Results and Analyses.} The results are shown in Figure \ref{fig:RQ2_results}. The horizontal and vertical axes represent the number of erroneous and correct programs, respectively. The stars in Figure \ref{fig:RQ2_results} denote the results of the first control group. The curves denote the results of the second and third groups, respectively. Each curve reflects the results of an approach under different thresholds. We expect the LLMs to generate more correct programs and fewer erroneous programs. Thus, the closer a curve is to the upper left corner, the better the approach's performance. 

\textbf{\method effectively decreases the number of erroneous programs shown to developers.} Compared to the results of the first group, \method significantly decreases the number of erroneous programs. However, \method and baselines lose a few correct programs.

\textbf{\method outputs more correct programs and fewer erroneous programs than baselines under different thresholds.} As shown in Figure \ref{fig:RQ2_results}, the curves of \method are closer to the upper left corners. With the same number of erroneous programs, \method can show more correct programs. Similarly, with the same number of correct programs, \method can show fewer erroneous programs. The results show that \method brings stable improvements over baselines under different thresholds.

\greyboxb{Answer to RQ2: } {\method effectively decreases the number of erroneous programs shown to developers. Besides, \method outperforms baselines by outputting more correct programs and fewer erroneous programs.}

\subsection{RQ3: Effectiveness of our multi-modal estimator}
\label{sec:results:RQ3_estimator}

As stated in Section \ref{sec:method:confidence}, a critical stage in \method is to estimate the LLMs' confidence by measuring the similarity between the generated programs. We propose to measure the similarity from multiple modalities. In the RQ3, we design and compare eight other estimators to our estimator to validate its effectiveness. We also analyze the contributions of different modalities.

\begin{table*}[t]
\centering
\caption{The performance of different estimators on \bench (python).}
\vspace{-0.3cm}
\label{tab:RQ3_python}
\resizebox{\linewidth}{!}{
\begin{tabular}{lcccccccc}
\toprule
\multirow{2}{*}{Estimator} & \multicolumn{2}{c}{CodeLlama-13B-Ins} & \multicolumn{2}{c}{CodeLlama-7B-Ins} & \multicolumn{2}{c}{DS-Coder-6.7B-Ins} & \multicolumn{2}{c}{DS-Coder-1.3B-Ins} \\
 & AUROC & AUCPR & AUROC & AUCPR & AUROC & AUCPR & AUROC & AUCPR \\ \midrule
BLEU & 71.55\% & 52.91\% & 72.15\% & 44.95\% & 62.35\% & 52.47\% & 72.58\% & 49.85\% \\
CodeBLEU & 72.07\% & 53.54\% & 72.88\% & 45.92\% & 62.54\% & 53.48\% & 71.66\% & 49.32\% \\
Edit Similarity & 72.64\% & 52.39\% & 70.42\% & 43.83\% & 60.19\% & 49.46\% & 70.31\% & 45.65\% \\
Embedding Similarity & 69.40\% & 49.00\% & 67.46\% & 39.42\% & 62.04\% & 52.91\% & 68.88\% & 46.70\% \\ \midrule
Ours w/o Text & 72.74\% & 53.12\% & 71.20\% & 44.13\% & 62.04\% & 52.91\% & 71.31\% & 47.18\% \\
Ours w/o Syntactic & 72.15\% & 51.77\% & 72.99\% & 45.84\% & 62.04\% & 52.91\% & 73.04\% & 50.30\% \\
Ours w/o Dataflow & 73.59\% & 54.27\% & 72.99\% & 45.84\% & 62.04\% & 52.91\% & 71.96\% & 48.78\% \\
Ours w/o Embedding & 71.07\% & 51.85\% & 72.15\% & 44.95\% & 63.53\% & 54.31\% & 71.48\% & 48.63\% \\ \midrule
\rowcolor[rgb]{ .741,  .843,  .933}
Ours & \textbf{74.05\%} & \textbf{54.83\%} & \textbf{73.91\%} & \textbf{46.55\%} & \textbf{63.73\%} & \textbf{54.38\%} & \textbf{73.36\%} & \textbf{50.96\%} \\
\bottomrule
\end{tabular}}
\vspace{-0.3cm}
\end{table*}

\begin{table*}[t]
\centering
\caption{The performance of different estimators on \bench (java).}
\vspace{-0.3cm}
\label{tab:RQ3_java}
\resizebox{\linewidth}{!}{
\begin{tabular}{lcccccccc}
\toprule
\multirow{2}{*}{Estimator} & \multicolumn{2}{c}{CodeLlama-13B-Ins} & \multicolumn{2}{c}{CodeLlama-7B-Ins} & \multicolumn{2}{c}{DS-Coder-6.7B-Ins} & \multicolumn{2}{c}{DS-Coder-1.3B-Ins} \\
 & AUROC & AUCPR & AUROC & AUCPR & AUROC & AUCPR & AUROC & AUCPR \\ \midrule
BLEU & 66.46\% & 51.53\% & 69.79\% & 44.18\% & 63.84\% & 63.29\% & 67.36\% & 57.08\% \\
CodeBLEU & 66.86\% & 51.92\% & 70.14\% & 45.55\% & 63.84\% & 64.17\% & 67.49\% & 57.64\% \\
Edit Similarity & 66.41\% & 51.55\% & 69.09\% & 44.42\% & 63.06\% & 62.16\% & 67.83\% & 57.89\% \\
Embedding Similarity & 55.32\% & 37.91\% & 62.39\% & 32.39\% & 54.06\% & 55.49\% & 62.96\% & 55.86\% \\ \midrule
Ours w/o Text & 66.84\% & 51.13\% & 69.33\% & 45.87\% & 62.77\% & 64.05\% & 68.12\% & 57.48\% \\
Ours w/o Syntactic & 68.12\% & 53.48\% & 69.63\% & 45.77\% & 63.36\% & 63.45\% & 68.20\% & 59.90\% \\
Ours w/o Dataflow & 67.97\% & 52.81\% & 70.51\% & 45.95\% & 63.86\% & 63.85\% & 68.25\% & 57.33\% \\
Ours w/o Embedding & 67.90\% & 52.55\% & 70.30\% & 46.12\% & 64.60\% & 64.06\% & 68.08\% & 57.73\% \\ \midrule
\rowcolor[rgb]{ .741,  .843,  .933}
Ours & \textbf{68.46\%} & \textbf{53.95\%} & \textbf{71.73\%} & \textbf{46.73\%} & \textbf{64.71\%} & \textbf{64.17\%} & \textbf{68.21\%} & \textbf{59.92\%} \\
\bottomrule
\end{tabular}}
\vspace{-0.2cm}
\end{table*}

\textbf{Settings.} The critical point of an estimator is the code similarity metric. We collect some popular metrics and design the following estimators for comparison:

\textbf{\ding{182} BLEU-based Estimator.} It uses the BLEU \cite{BLEU} as the similarity metric. BLEU is a classic text similarity metric widely used in code generation \cite{SkCoder,TreeGen,TRANX}. BLEU computes the $n$-gram similarity between programs:
\begin{equation}
\text{BLEU}= \text{BP} \cdot \exp \left(\sum_{n=1}^{4} \frac{1}{4} \log p_{n}\right)
\end{equation}
where $p_n$ is the ratio of overlapping subsequences of length $n$ between two programs. 

\textbf{\ding{183} CodeBLEU-based Estimator.} It uses the CodeBLEU \cite{CodeBLEU} as the similarity metric. CodeBLEU is a variant of BLEU and is designed for code. CodeBLEU revises the weights in BLEU and adds the similarities in code structures. The details of CodeBLEU can be found in its original paper \cite{CodeBLEU}.

\textbf{\ding{184} Edit Similarity-based Estimator.} It uses the edit similarity as the similarity metric. Edit similarity is computed based on the edit distance \cite{Lev_Distance}. The edit distance between two strings is the minimum number of edit operations (\ie insertions, deletions, or substitutions) required to transform one string into the other. We compute the edit similarity as follows:
\begin{equation}
\text{Edit Similarity} = 1 - \frac{Lev(c_i, c_j)}{max(len(c_i), len(c_j))}
\end{equation}
where $c_i$ and $c_j$ are two programs. $Lev(\cdot)$ is to compute token-level edit distances between programs.
    
\textbf{\ding{185} Embedding Similarity-based Estimator.} It considers the embedding similarity as the similarity metric. Following our estimator, it leverages an LLM (\ie DeepSeek Coder-Instruct-6.7B) to convert programs into dense embedding and outputs the cosine similarity between embeddings.

Besides the above four estimators, we make four variants of our multi-modal estimators by removing different modalities independently. For example, \ding{186} Ours w/o Text Similarity removes the text similarity. Similarly, we make \ding{187} Ours w/o Syntactic Similarity, \ding{188} Ours w/o Dataflow Similarity, and \ding{189} Ours w/o Embedding Similarity.

\textbf{Results and Analyses.} Table \ref{tab:RQ3_python} and \ref{tab:RQ3_java} shows the performance of \method with different estimators. ``Ours'' means the proposed multi-modal estimator. 

\textbf{Our estimator performs better than other design choices.} As shown in Table \ref{tab:RQ3_python} and \ref{tab:RQ3_java}, our estimator outperforms other designs on four LLMs and two programming languages. We attribute the improvements to our proposed multi-modal code similarity measurement. It offers a comprehensive view to measure the similarities between LLM-generated programs and is beneficial to estimating the certainty of LLMs. In the future, practitioners can introduce more advanced code similarity metrics and further enhance \method.

\textbf{All four modalities contribute to the performance of \method.} As shown in Table \ref{tab:RQ3_python} and \ref{tab:RQ3_java}, removing any of the modalities will result in a degradation of \method's performance. The results validate our motivation that the similarities between programs are reflected in different aspects. Thus, it is necessary to measure the code similarity from multiple modalities. 

\greyboxb{Answer to RQ3: } {Our multi-modal estimator performs better than the other eight estimators. Besides, an ablation study shows that all four modalities contribute to the performance of \method.}
\vspace{-0.2cm}
\section{Discussion}
\label{sec:discussion}

\begin{figure}[htp]
\centering
\includegraphics[width=0.9\linewidth]{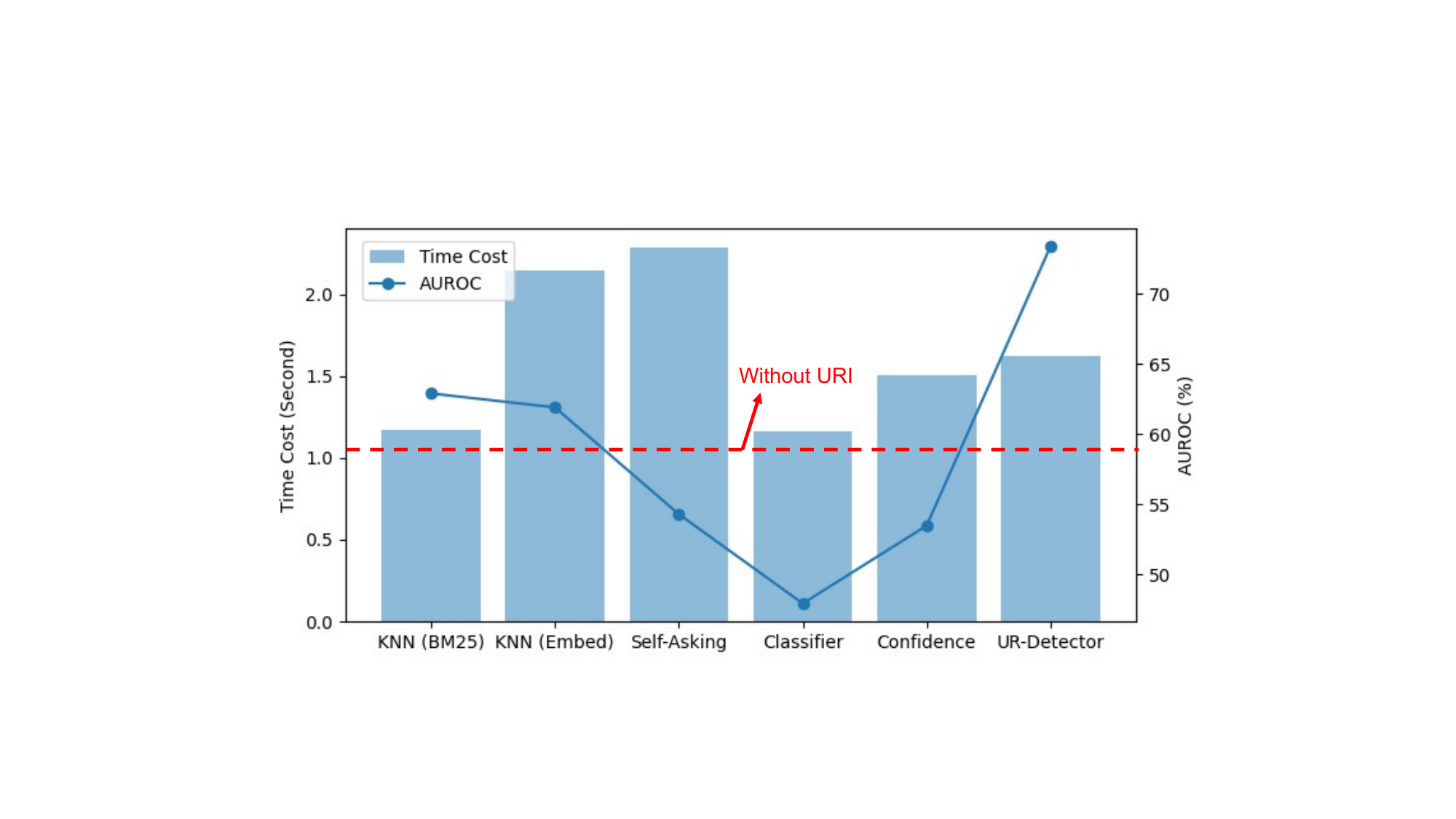}
\vspace{-0.2cm}
\caption{The comparison of different approaches regarding time costs and AUROC. The red dashed line shows the time cost when showing the generated code indiscriminately. ``Ours ($N$)`` means \method with a sampling size $N$. \method adds slight time overheads (approximately 0.4 seconds) and performs better than all baselines in AUROC.}
\label{fig:time_efficiency}
\vspace{-0.4cm}
\end{figure}

\subsection{Time Efficiency of \method}
\label{sec:discussion:efficiency}

Code generation tools typically require outputting programs efficiently. Thus, it is necessary to measure the time efficiency of \method. We select DS-Coder-1.3B-Ins as the base model and apply \method to it. Then, we calculate the average time costs of \method to process a requirement in \bench (python). It includes the time costs for estimating LLMs' confidence and outputting programs. We evaluate baselines in the same way. We run this experiment on an NVIDIA A100 GPU with 40GB memory. 

Figure \ref{fig:time_efficiency} shows the time costs (seconds) and AUROC (\%) of different approaches. ``Ours ($N$)`` means our \method with a sampling size $N$. The red dashed line shows generating and showing the code indiscriminately. \method adds slight time overheads (approximately 0.4 seconds) and performs better than all baselines in AUROC. Because the steps (\eg code generation and confidence estimation) in \method can be parallelized, a larger $N$ does not result in more time overheads but requires more memory. We set the default $N$ to 20 because a larger $N$ slightly improves AUROC. We recommend that practitioners set the sampling size according to specific application scenarios.

\vspace{-0.2cm}
\subsection{Threats To Validity}
\label{sec:discussion:threats}

There are two main threats to the validity of our work.

\textbf{The impact of inference settings.} In Section 
\ref{sec:benchmark:collection}, we annotate requirements as passed and failed requirements based on whether LLMs can generate the correct programs for them. In practice, the performance of LLMs on code generation is sensitive to inference settings, \eg sampling algorithms \cite{AdapT} and prompt designs \cite{AceCoder,SCoT}. To address this threat, we design a default inference setting in the label annotation. The inference setting is designed based on an interview of ten experienced developers and is described in Section \ref{sec:benchmark:collection}. Furthermore, our data collection pipeline and \method do not rely on specific inference settings. Practitioners can easily apply our work to different inference settings.

\textbf{Comparison with test case generation.} Recent studies \cite{Test_Gen,CodaMosa} have proposed automatic test case generation. Intuitively, we can use the generated test cases to check LLM-generated programs and only show passed programs. However, we think this approach has the following limitations.
\begin{itemize}[leftmargin=*]
    \item \textbf{Unreliable.} Auto-generated test cases may be untrustworthy \cite{Test_Gen}. Even if a program passes the generated test cases, it could still be wrong. 
    \item \textbf{Heavy.} Executing test cases requiring installing all dependencies properly. This is especially problematic when developers want code suggestions in a project with multiple files and dependencies. Even if these dependencies are satisfied, many realistic programming scenarios involve incomplete code under active development where execution is infeasible.
    \item \textbf{Unsafe.} LLM-generated programs may have security risks (\eg deleting files). Thus, executing such programs requires heavy-weight isolation mechanisms such as a sandbox environment. 
\end{itemize}
In comparison, our \method addresses the above limitations. \method provides a new idea to measure the correctness of programs by estimating the confidence of LLMs. Extensive experiments in four LLMs have shown the effectiveness of our idea. Besides, \method is execution-free and does not rely on any dependencies. Developers can easily use \method on different LLMs and code projects.

\vspace{-0.2cm}
\section{Related Work}
\label{sec:related_work}

To our knowledge, this paper is the first to explore selectively showing LLM-generated programs to developers. In this section, we summarize related studies with motivations similar to this paper. These related studies can be divided into two groups, including code confidence estimation and requirement confidence estimation. We describe their details as follows.

\textbf{Code confidence estimation} aims to estimate the LLMs' confidence in the generated programs. The confidence reflects the reliability of the generated programs. A recent work \cite{Code_Confidence} studied a few approaches to estimating code confidence, \eg average token probability and self-asking. The details of these approaches are in Section \ref{sec:study_design:baselines}. The studied approaches are initially designed for natural language generation and have sub-optimal performance in code generation. 

\textbf{Requirement confidence estimation} is to estimate the LLMs' confidence in requirements. For low-confidence requirements, LLMs should refuse to answer. No public studies are exploring this topic on code generation. We draw on related work \cite{Self-Guided-RAG} in question answering and apply three effective approaches to code generation, including self-asking, K-nearest neighbor search, and requirement classifier. The details of these approaches are described in Section \ref{sec:study_design:baselines}.

\textbf{Comparison with this paper.} The differences between the above studies and this paper are two-fold. First, this paper focuses on a new practical problem (\ie selectively showing LLMs-generated programs). To address this problem, we propose a new approach - \method and release a new benchmark. Second, our proposed \method provides a new idea to estimate the LLMs' confidence, \ie measuring the similarity between LLM-generated multiple programs. In the RQ1 (Section \ref{sec:results:RQ1}) and RQ2 (Section \ref{sec:results:RQ2_reducing}), the confidence computed by \method substantially outperforms the above work on four LLMs.

\vspace{-0.2cm}
\section{Conclusion and Future Work}
\label{sec:conclusion}

LLMs may generate erroneous programs, and we should avoid showing them to developers. To achieve this goal, we propose a novel code generation approach named \method. It selectively shows LLM-generated programs to developers based on the LLMs' confidence. We propose an effective approach to estimate the LLMs' confidence in code generation. The confidence provides valuable insights into the reliability of LLM-generated programs. We collect a multilingual benchmark and evaluate \method on it. Extensive experiments show the effectiveness of \method in estimating LLMs' confidence and reducing the shown erroneous programs.

\textbf{Future Work.} We summarize the potential directions of our work to facilitate the application of LLMs in software development. 
\begin{itemize}[leftmargin=*]
    \item \textbf{Exploring more accurate confidence estimation approaches.} Accurate confidence estimation is valuable in explaining the LLMs' behavior. In the future, researchers can explore more accurate confidence estimation approaches. For example, we can estimate the LLMs' confidence based on their internal hidden states and attention scores. Besides, we can also teach LLMs to verbalize their confidence by instruction tuning.
    \item \textbf{Designing more advanced generation techniques.} \method refusal to show programs when LLMs are uncertain. In the future, we will explore alternatives. For example, we retrieve relevant programs from a codebase and re-generate programs based on retrieved results. Besides, we can teach LLMs to verbalize why they are uncertain. Then, developers can provide more information to address the LLMs' uncertainty. 
    \item \textbf{Motivating discussion in other fields.} The research problem in this paper probably exists in other code-related tasks, \eg code repair and vulnerability detection. Our work will motivate broad discussions in these tasks. For example, on vulnerability detection, we can decrease the number of false positives in LLMs' predictions based on the LLMs' confidence. 
\end{itemize}



\bibliographystyle{ACM-Reference-Format}
\bibliography{reference}

\end{document}